\documentclass[english,aps,manuscript]{revtex4}
\usepackage[T1]{fontenc}
\usepackage[latin1]{inputenc}
\usepackage{amsmath}

\makeatletter
\newcommand{\bee}{\begin{equation}}
\newcommand{\ee}{\end{equation}}
\newcommand{\beea}{\begin{eqnarray}}
\newcommand{\eea}{\end{eqnarray}}

\usepackage{babel}
\makeatother

\begin{document}

\section*{COLO-HEP-540}

\begin{center}
\textbf{\Large Heterotic Strings on Generalized Calabi-Yau Manifolds
and Kaehler Moduli Stabilization}
\par\end{center}{\Large \par}

\begin{center}
\vspace{0.3cm}
\par\end{center}

\begin{center}
{\large S. P. de Alwis$^{\dagger}$ }
\par\end{center}{\large \par}

\begin{center}
Physics Department, University of Colorado, \\
 Boulder, CO 80309 USA
\par\end{center}

\begin{center}
\vspace{0.3cm}
\par\end{center}

\begin{center}
\textbf{Abstract}
\par\end{center}

\begin{center}
\vspace{0.3cm}
\par\end{center}

Compactifications of heterotic string theory on Generalized Calabi-Yau
manifolds have been expected to give the same type of flexibility
that type IIB compactifications on Calabi-Yau orientifolds have. In
this note we generalize the work done on half-flat manifolds by other
authors, to show how flux quantization occurs in the general case,
by starting with a basis of harmonic forms and then extending it.
However it turns out that only the axions associated with the non-harmonic
directions in the space of Kaehler moduli, can be stabilized by the
geometric (torsion) terms. Also we argue that there are no supersymmetric
extrema of the potential when the second (and fourth) cohomology groups
on the manifold are non-trivial. We suggest that threshold corrections
to the classical gauge coupling function could solve these problems.

\vfill

$^{\dagger}$ {\small e-mail: dealwiss@colorado.edu}{\small \par}

\eject

\section{Introduction}

Most of the recent work on flux compactifications of string theory
has focused on type IIB on Calabi-Yau (CY) orientifolds (see \citep{Grana:2005jc,Douglas:2006es}
for reviews). However it is far more difficult to get the standard
model (or the MSSM) there than it is to get it in Heterotic compactifications.
The principal reason is chirality which in Heterotic theory can actually
be obtained quite easily, while in type IIB this requires somewhat
elaborate constructions. Getting a Grand Unified Theory (GUT) is even
more difficult in the IIB context.

On the other hand in type IIB on CY orientifolds one can turn on two
types of fluxes; NS-NS (H-flux) and RR to stabilize (after including
also non-perturbative terms) all (closed string) moduli as well as
the dilaton. The existence of two types of fluxes gives enough fine-tuning
freedom, that it is easy to see that there would be many models that
can have the dilaton stabilized at an acceptable value, a gravitino
mass many orders of magnitude below the Planck scale (if one wants
low energy supersymmetry (SUSY)), and a zero (or highly suppressed
value) for the cosmological constant (CC).

However in the Heterotic string compactified on CY one can only turn
on one type of flux, namely H-flux, and even with non-perturbative
terms (which are in fact necessary to stabilize the dilaton) one cannot
stabilize the Kaehler moduli - in fact one has a no-scale model. Furthermore,
typically the dilaton will be stabilized in the strong coupling regime.
One can introduce threshold corrections to these non-perturbative
corrections to stabilize the Kaehler moduli, but then of course one
needs to tune the cosmological constant to an acceptable value and
it is not clear that one has enough freedom to do so.

It has been suggested that these problems can be overcome by compactifying
the heterotic string on generalized Calabi-Yau manifolds \citep{deCarlos:2005kh}\citep{Benmachiche:2008ma}
(for earlier work see \citep{Cardoso:2002hd,LopesCardoso:2003sp,Becker:2003yv,Becker:2003sh,Gurrieri:2004dt,Frey:2005zz}).
Much of this work has been in the context of so-called half-flat manifolds.
In constructing the potential for the moduli in these works a basis
of forms is introduced that is not harmonic, and it is not clear how
this could still result in a potential for the moduli that is determined
in terms of integer flux parameters. Nevertheless as shown in \citep{deCarlos:2005kh}
this quantization condition still holds for all but one parameter.
However even this appeared to be difficult to establish for generalized
(non-half-flat) CY manifolds. In this paper we show that by a judicious
choice of basis forms it is possible to show the quantization of the
flux parameters, but it is still unclear how to show the quantization
of parameters corresponding to the torsion of the manifold. We also
show that if the cohomology groups $H^{2}/H^{4}$ are non-trivial
then there is no supersymmetric solution even for generalized CY manifolds.
What happens is that the SUSY equations for the corresponding Kaehler
moduli have no solution. There is no obstruction, as far as we can
see, to finding non-supersymmetric solutions, but even when there
are such stable points, the axions corresponding to those Kaehler
moduli will be flat directions in the potential just as in the case
of half-flat manifolds \citep{deCarlos:2005kh}.

\section{Constructing a Basis}

Let $\hat{Y}$ be a generalized Calabi-Yau (CY) manifold obtained
by deforming a Calabi-Yau manifold $Y$ whose cohomology groups $H^{3}(Y)$
and $H^{2}/H^{4}(Y)$ have dimensions $2b_{3}\equiv2(h_{12}+1)$ and
$b_{2}=b_{4}=h_{11}$ respectively. As with the underlying CY manifold
$Y$ there is a globally defined real two form $J$ and a globally
defined complex three form $\Omega$ in $\hat{Y}$, but unlike in
$Y$ on $\hat{Y}$, $dJ\ne0,$ and $d\Omega\ne0$. So the dimensions
of the cohomology groups $H^{3}(\hat{Y}),H^{2}/H^{4}(\hat{Y})$ are
such that $\hat{b_{3}}<b_{3},\,\hat{b_{2}}<b_{2}$. $ $ A basis of
harmonic forms for $H^{3}(\hat{Y})$ is\begin{equation}
\alpha_{a},\beta^{a};\, a=1,2,\ldots,\hat{b}_{3}\equiv\hat{h}_{21}+1,\label{eq:b3}\end{equation}
and for $H^{2}/H^{4}$ \begin{equation}
\omega_{i},\tilde{\omega^{i}};\, i=1,2,\ldots,\hat{b}_{2}=\hat{b}_{4}\equiv\hat{h}_{11}.\label{eq:b2b4}\end{equation}
These satisfy the usual relations for a harmonic basis,\begin{eqnarray}
\int_{\hat{Y}}\alpha_{a}\wedge\beta^{b} & =\int_{A^{b}}\alpha_{a}=\int_{B^{b}}\beta^{b}= & \delta_{a}^{b},\nonumber \\
\int_{\hat{Y}}\alpha_{a}\wedge\alpha_{b}=\int_{\hat{Y}}\beta^{a}\wedge\beta^{b} & =\int_{B_{b}}\alpha_{a}=\int_{A^{a}}\beta^{b}= & 0.\label{eq:3formnorm}\end{eqnarray}
Here $A^{a}(B_{a})$ are three cycles dual to $\beta^{a}(\alpha_{a})$.
Also \begin{equation}
\int_{\hat{Y}}\omega_{i}\wedge\tilde{\omega}^{j}=\int_{C_{2}^{j}}\omega_{i}=\int_{C_{4}^{i}}\tilde{\omega}^{j}=\delta_{i}^{j},\label{eq:2-4formnorm}\end{equation}
where $C_{2}^{j}(C_{4}^{i})$ are cycles dual to $\tilde{\omega}^{j}(\omega_{i})$.
Henceforth all integrals without a subscript will be taken over the
whole manifold $\hat{Y}$. Note that in this section we have chosen
units such that $2\pi\sqrt{\alpha'}=1$.

Now as remarked earlier, on $\hat{Y}$\begin{equation}
dJ\ne0,\, d\Omega\ne0.\label{eq:dJOmega}\end{equation}
Clearly because of (\ref{eq:dJOmega}) $J$ and $\Omega$ cannot be
expanded in the above harmonic basis. We need to append to this basis
additional two/four forms and three forms \[
\omega_{\kappa},\tilde{\omega}^{\kappa},\,\kappa=\hat{b}_{2}+1,\ldots,b_{2};\,\alpha_{\gamma},\beta^{\gamma},\,\gamma=\hat{b}_{3}+1,\ldots b_{3}.\]
These are neither closed nor co-closed and may be written as\begin{equation}
\omega_{\kappa}=\delta\omega_{\kappa(3)}+d\omega_{\kappa(2)}^{},\,\tilde{\omega}^{\kappa}=\delta\omega_{(5)}^{\kappa}+d\omega_{(3)}^{\kappa};\,\alpha_{\gamma}=\delta\alpha_{\gamma(4)}+d\alpha_{\gamma(2)},\,\beta^{\gamma}=\delta\beta_{(4)}^{\gamma}+d\beta_{(2)}^{\gamma}.\label{eq:243form0}\end{equation}
They can be taken to satisfy the relations\begin{eqnarray}
\int\alpha_{\gamma}\wedge\beta^{a} & = & \int\alpha_{\gamma}\wedge\alpha_{a}=\int\alpha_{a}\wedge\beta^{\gamma}=\int\beta^{\gamma}\wedge\beta^{a}=0,\,\int\alpha_{\gamma}\wedge\beta^{\delta}=\delta_{\gamma}^{\delta},\label{eq:0anorm}\\
\int\omega_{\kappa}\wedge\tilde{\omega^{i}} & = & \int\omega_{i}\wedge\tilde{\omega^{\kappa}}=0,\,\int\omega_{\kappa}\wedge\tilde{\omega}^{\lambda}=\delta_{\kappa}^{\lambda}.\label{eq:0inorm}\end{eqnarray}
Note that in the above we have made the usual assumption common in
the physics literature (see for example section 3 of \citep{Grana:2005ny})
that these generalized manifolds are closely related to the CY manifold
of which it is a generalization, so that there is a finite basis of
forms in which $J$ and $\Omega$ can be expanded in, that is related
to the harmonic basis of the corresponding CY. Combining equations
\eqref{eq:3formnorm}\eqref{eq:2-4formnorm}\eqref{eq:0anorm}\eqref{eq:0inorm}
we may write \begin{eqnarray}
\int\alpha_{A}\wedge\beta^{B} & = & \delta_{A}^{B},\,\int\alpha_{A}\wedge\alpha_{B}=\int\beta^{A}\wedge\beta^{B}=0,\, A,B=1,\ldots b_{3},\label{eq:ABnorm}\\
\int\omega_{I}\wedge\tilde{\omega^{J}} & = & \delta_{I}^{J},\, I,J=1,\ldots b_{2}.\label{eq:IJnorm}\end{eqnarray}
Note however there is no analog of the integrals over 3, 2 and 4-cycles
as in equations\eqref{eq:3formnorm}\eqref{eq:2-4formnorm} for these
additional forms which are not harmonic. 

Expanding the globally defined complex three form $\Omega$ in terms
of the basis three forms we have,\begin{equation}
\Omega=Z^{A}\alpha_{A}-G_{A}\beta^{A},\label{eq:Omegaexansion}\end{equation}
where in the related CY manifold $Y$ the $Z^{A}$ would be homogeneous
(complex) coordinates on the space of complex structures and $G_{A}=\frac{\partial G}{\partial Z^{A}},$
with $G(Z^{A})$ being a homogeneous function of degree two. 

Similarly the globally defined two form $J$ may be expanded in our
basis two forms as\begin{equation}
J=t^{I}\omega_{I}=t^{i}\omega_{i}+t^{\kappa}\omega_{\kappa},\label{eq:J}\end{equation}
where $t^{I}$ would be the set of Kaehler moduli of the underlying
CY manifold $Y$. Now we have, expanding in terms of our basis three
forms,\[
d\omega_{\kappa}=p_{\kappa A}\beta^{A}-q_{\kappa}^{A}\alpha_{A}.\]
Using the ortho-normality relations (\ref{eq:ABnorm}) \eqref{eq:IJnorm}
and the harmonicity of $\alpha_{a},\beta^{a}$ we have \[
p_{\kappa a}=\int_{\hat{Y}}\alpha_{a}\wedge d\omega_{\kappa}=0,\, q_{\kappa}^{a}=\int_{\hat{Y}}\beta^{a}\wedge d\omega_{\kappa}=0.\]
So \begin{equation}
d\omega_{\kappa}=p_{\kappa\gamma}\beta^{\gamma}-q_{\kappa}^{\gamma}\alpha_{\gamma}.\label{eq:domega0}\end{equation}
Thus we get \begin{equation}
dJ=t^{I}d\omega_{I}=t^{\kappa}d\omega_{\kappa}=\sum_{\kappa=\hat{b_{2}}+1}^{b_{2}}t^{\kappa}(p_{\kappa\gamma}\beta^{\gamma}-q_{\kappa}^{\gamma}\alpha_{\gamma}).\label{eq:dJ}\end{equation}

\section{Quantization of H-flux}

Now we can expand $H$ flux in terms of the basis three forms %
\footnote{The normalization factor in the equation here is a consequence of
taking the convention used in \citep{deCarlos:2005kh}.%
}:\begin{equation}
H_{flux}=\frac{1}{6\sqrt{8}}(\mu^{A}\alpha_{A}-\epsilon_{A}\beta^{A})\label{eq:Hflux}\end{equation}
If we work to lowest order in the $\alpha'$ expansion and ignore
the Green-Schwarz (GS) term (or take the standard embedding of the
tangent bundle connection in the gauge connection) we have $dH_{flux}=0$.
The flux must also satisfy the equation of motion $d*H_{flux}=0$
(again ignoring the GS term) i.e. $H_{flux}$ is harmonic. So the
expansion in (\ref{eq:Hflux}) runs only over the harmonic part of
the basis:\begin{equation}
H_{flux}=\frac{1}{6\sqrt{8}}(\mu^{a}\alpha_{a}-\epsilon_{a}\beta^{a})\label{eq:Hfluxharm}\end{equation}
\[
\]

Now since $\alpha_{a},\beta^{a}$ are harmonic we have (see \eqref{eq:3formnorm})
after restoring the string scale,\begin{equation}
\int_{A^{b}}\alpha_{a}=\int_{B_{a}}\beta^{b}=(2\pi\sqrt{\alpha'})^{3}\delta_{b}^{b}.\label{eq:intalphabetacycle}\end{equation}
Using the relations \eqref{eq:ABnorm} (which now acquire a factor
$(2\pi\sqrt{\alpha'})^{6}$ on the RHS) we have (since $dH=0$)\begin{equation}
(2\pi\sqrt{\alpha'})^{3}\int_{A^{a}}H=\int H\wedge\beta^{a}=\frac{1}{6\sqrt{8}}\mu^{a}(2\pi\sqrt{\alpha'})^{6}.\label{eq:Halpha}\end{equation}
 The first equation above is Poincare duality and the second follows
from \eqref{eq:Hflux} and the orthonormality relations \eqref{eq:ABnorm}.
Quantization of H-flux (coming from it being sourced by strings) implies
\begin{equation}
\int_{C_{3}}H=(2\pi\sqrt{\alpha'})^{2}n,\, n\in{\cal Z}.\label{eq:Hquant}\end{equation}
Thus we have from \eqref{eq:Halpha} and a similar equation involving
an $B^{a}$ cycle and its associated dual form $\alpha_{a}$,\begin{equation}
\mu^{a}=\frac{6\sqrt{2}}{\pi\sqrt{\alpha'}}n_{A}^{a},\,\epsilon_{a}=\frac{6\sqrt{2}}{\pi\sqrt{\alpha'}}n_{Ba}^{};\, n_{A}^{a},n_{Ba}\in{\cal Z}.\label{eq:muepsilonquant}\end{equation}

\section{Stabilization of Kaehler moduli}

The Kaehler potential of the theory is given by \[
K=-\ln(S+\bar{S})-\ln(8{\cal K})-\ln(8\tilde{{\cal K}})\]
 where\begin{eqnarray}
{\cal K} & = & d_{IJK}t^{I}t^{J}t^{K}\label{eq:calK}\\
\tilde{{\cal K}} & =\sum_{1}^{h_{12}} & \tilde{d}_{abc}z^{a}z^{b}z^{c}\label{eq:tildecalK}\end{eqnarray}
 where $t^{I}$ are the Kaehler moduli and $z^{a}$ are the (real
parts of) the complex structure moduli and $d_{IJK},\tilde{d}_{abc}$
are intersection numbers for the 2 cycles on the manifold and its
mirror. The holomorphic coordinates on the two moduli spaces are $T=t^{i}+i\tau^{i}$
and $Z^{a}=z^{a}+i\zeta^{a}$. Also we write $S=s+i\sigma$ so that
the real fields $s,\sigma$ are respectively the model independent
dilaton and axion of the heterotic string. 

The GTVW superpotential of this theory is (see \citep{deCarlos:2005kh}
and references therein)\begin{equation}
{\cal W}=\int\Omega\wedge(H+idJ)\label{eq:GTVW}\end{equation}
From \eqref{eq:Omegaexansion} and \eqref{eq:dJ} we have \[
\int\Omega\wedge idJ=i(Z^{\gamma}p_{\kappa\gamma}-G_{\gamma}(Z)q_{\kappa}^{\gamma})t^{\kappa}\]
The field strength $H=dB+H_{flux}$, where we may expand the two form
field in terms of our basis two forms $B=-\tau^{I}\omega_{I}$, where
the $\tau^{I}$ are the model dependent axions of the underlying string
theory. So ignoring the 4D derivative of $\tau$ (which does not contribute
to the superpotential) we have\begin{equation}
H=-\tau^{\kappa}d\omega_{\kappa}+H_{flux}=-\tau^{\kappa}(p_{\kappa\gamma}\beta^{\gamma}-q_{\kappa}^{\gamma}\alpha_{\gamma})+H_{flux},\label{eq:Hexpansion}\end{equation}
where we've used $d\omega_{i}=0$ and \eqref{eq:domega0}. From \eqref{eq:Omegaexansion}
and \eqref{eq:Hexpansion} we have (using  \eqref{eq:ABnorm}) the
result\[
\int\Omega\wedge H=\frac{1}{6\sqrt{8}}(-Z^{a}\epsilon_{a}+G_{a}\mu^{a})-(Z^{\gamma}p_{\kappa\gamma}-G_{\gamma}(Z)q_{\kappa}^{\gamma})\tau^{\kappa}.\]
Thus the GTVW superpotential is given by\begin{eqnarray*}
{\cal W} & = & \frac{1}{6\sqrt{8}}(-Z^{a}\epsilon_{a}+G_{a}\mu^{a})+i(Z^{\gamma}p_{\kappa\gamma}-G_{\gamma}(Z)q_{\kappa}^{\gamma})T^{\kappa}.\\
 & = & Z^{A}P_{A}-G_{A}Q^{A},\end{eqnarray*}
where \[
P_{A}=(ip_{\kappa\gamma}T^{\kappa}\delta_{A}^{\gamma}-\frac{\epsilon_{a}}{6\sqrt{8}}\delta_{A}^{a}),\, Q^{A}=(iq_{\kappa}^{\gamma}T^{\kappa}\delta_{\gamma}^{A}-\frac{\mu^{a}}{6\sqrt{8}}\delta_{a}^{A}).\]

Of course with this superpotential the dilaton will not be stabilized.
So we need to add a non-perturbative (NP) term coming from gaugino
condensation. Typically in the $E_{8}\times E_{8}$ theory the condensing
gauge group is taken to be the hidden sector $E_{8}$ or a subgroup
thereof. Thus the total superpotential is \begin{equation}
W={\cal W}+ke^{-cS},\label{eq:Wtotal}\end{equation}
where c is a group theoretic number which depends on the gauge group
and its matter representations. The pre-factor $k$ is a holomorphic
function of the moduli if we take threshold corrections into account.
However for the moment let us ignore this and take $k$ to be a constant
as is usually done. Let us look at the Kaehler derivatives for the
moduli.\begin{eqnarray}
D_{S}W & = & -kce^{-cS}-\frac{1}{S+\bar{S}}W\label{eq:DSW}\\
D_{A}W & = & P_{A}-G_{AB}Q^{B}-\frac{\partial_{A}\tilde{{\cal K}}}{\tilde{{\cal K}}}W\label{eq:DaW}\\
D_{\kappa}W & = & i(p_{\kappa\gamma}Z^{\gamma}-G_{\gamma}q_{\kappa}^{\gamma})-\frac{\partial_{\kappa}{\cal K}}{{\cal K}}W\label{eq:D0W}\\
D_{i}W & = & -\frac{\partial_{i}{\cal K}}{{\cal K}}W\label{eq:DiW}\end{eqnarray}
These expressions tell us that there is no supersymmetric solution
( i.e. a solution of $D_{\Phi}W=0$ for all moduli) if the cohomology
group $H^{2}(\hat{Y})$ is non-trivial. The only way to satisfy the
equation $D_{T^{i}}W=0$ is to have $t^{i}\rightarrow\infty$. Note
that the option of tuning $W$ to zero is not really available since
if that were the case $S\rightarrow\infty$ i.e. the theory would
be at zero coupling. Thus it appears that generalized CY manifolds
do not solve Strominger's original problem \citep{Strominger:1986uh}
of finding supersymmetric solutions to the heterotic string in the
presence of fluxes unless the manifold has no harmonic two (or four)
forms and hence no two (or four) cycles. Thus supersymmetry is necessarily
broken as was the case for compactification on CY manifolds. Of course
if there are no harmonic two forms, the last equation does not exist
and we can have a supersymmetric solution.

We can avoid this conclusion if we include threshold corrections to
the classical gauge coupling function $S$ which will result in the
prefactor $k$ of the NP term acquiring a dependence on the moduli
of the internal manifold %
\footnote{These threshold corrections have only been calculated for orbifolds
and typically in that case they are proportional to Dedekind $\eta$
functions of the moduli. %
}. In this case there are extra terms on the RHS of \eqref{eq:DaW}\eqref{eq:D0W}
and \eqref{eq:DiW}. In particular there would be a term $\partial_{T^{i}}ke^{-cS}$
on the RHS of \eqref{eq:DiW} so that we may have a supersymmetric
solution at finite values of all the Kaehler moduli.

There does not appear to be any obstruction to finding non-supersymmetric
minima of the potential with more than one Kaehler modulus. However
if $k$ is constant, then the axions corresponding to the moduli which
do not appear in $W$ (i.e. the ones corresponding to the harmonic
two and four forms), will not be stabilized.

\section{Acknowledgements}

I wish to thank Andrei Micu for very useful discussions on the contents
of this paper and for pointing out that a construction in an earlier
version was equivalent to the half-flat case. I also wish to thank
Jan Louis for introducing me to the subject, Siye Wu for discussions
on the mathematical approach to these manifolds, and the organizers
of the CERN string phenomenology institute where this work was started.
This research is supported in part by the United States Department
of Energy under grant DE-FG02-91-ER-40672.

\bibliographystyle{apsrev}
\bibliography{myrefs}

\end{document}